\documentclass[10pt,conference]{IEEEtran}

\usepackage[ left=0.625in, top=0.75in, right=0.75in, bottom=1.03in, headsep=0.25in, headheight=20pt, heightrounded, letterpaper] {geometry}

\usepackage[utf8]{inputenc}
\usepackage{lipsum}
\usepackage{comment}
\usepackage{xfrac}  

\usepackage{lipsum,microtype}
\usepackage{authblk}
\usepackage[ruled,vlined]{algorithm2e}
\usepackage{amsmath, nccmath}
\usepackage{geometry}
\usepackage{tabularx}
\usepackage{booktabs}
\usepackage[labelfont=bf,format=plain,justification=raggedright,singlelinecheck=true]{caption}
\usepackage{mathtools}
\usepackage{amssymb}
\usepackage{caption}
\usepackage{xcolor}
\usepackage{subcaption}
\usepackage{ulem}
\usepackage{graphicx}
\graphicspath{{./Fig/}} % Set the path to the "Fig" folder

\usepackage[ruled,vlined]{algorithm2e}
\usepackage{cite}
\newcounter{mycounter}
\usepackage{subfig} % For subfigures
\usepackage{fancyhdr}
\usepackage{textcomp}
\usepackage{blindtext}
\usepackage{optidef}
\title{Self-Play Ensemble Q-learning enabled Resource Allocation for Network Slicing}

\author[1]{Shavbo Salehi, Student Member, IEEE}
\author[1]{Pedro Enrique Iturria-Rivera}
\author[2]{Medhat Elsayed}
\author[2]{\\Majid Bavand}
\author[3]{Raimundas Gaigalas}
\author[2]{Yigit Ozcan}
\author[1]{Melike Erol-Kantarci, Senior Member, IEEE}
\affil[1]{School of Electrical Engineering and Computer Science, University of Ottawa, Ottawa, Canada}
\affil[2]{Ericsson Canada Inc., Ottawa, Canada}
\affil[3]{Ericsson AB, Stockhom, Sweden}%{Ericsson Inc., Stockholm, Sweden}
\affil[ ]{{Emails: \{ssale038, pitur008, melike.erolkantarci}\}@uottawa.ca}
\affil[ ]{{\{medhat.elsayed,majid.bavand, raimundas.gaigalas,yigit.ozcan}\}@ericsson.com}

\date{}
\begin{document}

\twocolumn

\maketitle
\begin{abstract}
In 5G networks, network slicing has emerged as a pivotal paradigm to address diverse user demands and service requirements. To meet the requirements, reinforcement learning (RL) algorithms have been utilized widely, but this method has the problem of overestimation and exploration-exploitation trade-offs. To tackle these problems, this paper explores the application of self-play ensemble Q-learning, an extended version of the RL-based technique. Self-play ensemble Q-learning utilizes multiple Q-tables with various exploration-exploitation rates leading to different observations for choosing the most suitable action for each state. Moreover, through self-play, each model endeavors to enhance its performance compared to its previous iterations, boosting system efficiency, and decreasing the effect of overestimation. 
For performance evaluation, we consider three RL-based algorithms; self-play ensemble Q-learning, double Q-learning, and Q-learning, and compare their performance under different network traffic. Through simulations, we demonstrate the effectiveness of self-play ensemble Q-learning in meeting the diverse demands within 21.92\% in latency, 24.22\% in throughput, and 23.63\% in packet drop rate in comparison with the baseline methods. Furthermore, we evaluate the robustness of self-play ensemble Q-learning and double Q-learning in situations where one of the Q-tables is affected by a malicious user. Our results depicted that the self-play ensemble Q-learning method is more robust against adversarial users and prevents a noticeable drop in system performance, mitigating the impact of users manipulating policies.

\end{abstract}


\section{Introduction}
\label{section:intro}

Network slicing stands as a key feature in the 5G networks domain, offering an effective and efficient approach to addressing various services' specific demands \cite{babbar2022role}. In this context, ultra-reliable low latency communication (URLLC), enhanced mobile broadband (eMBB), and massive machine-type communications (mMTC) are fundamental use cases \cite{liu2023network} which demand low-latency, high throughput, and supporting a massive number of connected devices, respectively. Using the network slicing strategy, 5G networks can address the requirements of URLLC, eMBB, and mMTC seamlessly in various slices \cite{liu2023network}. As a means of optimizing these slices' performance, machine learning-based (ML-based) algorithms, in particular reinforcement learning (RL), present promising solutions. Q-learning, as a type of RL, empowers algorithms to make intelligent decisions by learning from interactions with the environment. Furthermore Q-learning adapts to dynamic network conditions acquires optimal strategies over time, and provides efficient resource allocation \cite{zhou2021ran}. 

While the RL-based methods have a considerable effect on meeting the requirements of network slicing, they face some challenges such as overestimation bias, slow learning, complexity in large-scale network scenarios, and the need for extensive training data \cite{elsayed2019ai}. %\noteblue{Similarly, despite the robustness of Q-learning, resource allocation in 5G networks faces challenges \cite{xiong2019deep} due to the exploration-exploitation trade-off, hindering the algorithm in achieving optimal allocations for diverse slices \cite{alam2023q}.}
%Similarly, despite Q-learning robustness, it poses challenges when it comes to resource allocation in 5G networks \cite{xiong2019deep}. 
%\notered{Is there any citation that can support this statement? }. 
In determining optimal resource allocations for diverse slices, the RL algorithm encounters difficulty with the exploration-exploitation trade-off \cite{alam2023q}. The high-dimensional state spaces inherent in network slicing, representing various service requirements and dynamic network conditions, also lead to the challenge of effectiveness in complex environments \cite{elsayed2019ai}. To address these weaknesses, appropriate modifications to Q-learning are necessary to increase the efficiency and adaptability of resource allocation \cite{wang2021adaptive}.

Addressing the challenges raised by Q-learning for resource allocation in network slicing, alternative approaches such as double Q-learning, deep Q-learning (DQN), and double DQN \cite{9899734,majumdar2023towards} emerge as promising solutions.
%\notered{Please, review the date of some algorithms. For ex. SARSA is from 1994, saying that is a promising algorithm might not best way to put it.}. 
By utilizing two separate value functions for action selection and evaluation, double Q-learning mitigates the algorithm's exploration-exploitation trade-off, thereby improving its ability to determine optimal resource allocations \cite{rezazadeh2021zero}. DQN uses neural networks to handle high-dimensional state spaces, which provides a scalable solution for network slicing \cite{chiang2022deep}. 
%\sout{Using these advanced techniques, it is possible to improve generalization across diverse slices and enhance adaptability in dynamic network environments} \notered{It's been said, the advantage of DQN over Q-Learning. One question that can be raised is why didn't you choose that one.}. 
Beyond these methods, ensemble Q-learning, which employs multiple Q-tables, is proposed to address overestimation bias and the handling of complex interactions in Q-learning \cite{peer2021ensemble}.
%\notered{Is ensemble Q learning from 2021?}\noteblue{I am not sure but most of the papers are proposed after that.}. 

While double Q-learning, DQN, and ensemble Q-learning significantly contribute to mitigating challenges in Q-learning, there remain unresolved issues related to efficient exploration and adaptability in dynamic environments and heterogeneous objectives optimization. Moreover, their susceptibility to adversarial users is related to their concentration on the current state and experience for action selection. To tackle these challenges, in this paper self-play ensemble Q-learning is proposed, offering a solution that uses diverse learning strategies through iterative interactions with the agent's previous versions, in the context of resource allocation. Different from existing algorithms such as double Q-learning and Q-learning according to our simulation results, self-play ensemble Q-learning provides more efficient, scalable, and adaptive allocation strategies that address the limitations of Q-learning when applied to evolving and complex scenarios. While RL-based methods have a considerable effect on improving network slicing performance \cite{zhou2022knowledge}, they are quite vulnerable to malicious users \cite{salehi2023adverserial}. In this paper, to evaluate the robustness of methods against malicious users, such as double Q-learning and self-play ensemble Q-learning, we considered a scenario in which one of the tables is affected by a malicious user who chooses the action that leads to the lowest Q-value. While the double Q-learning algorithm performance drops considerably, self-play ensemble Q-learning, due to the agent's interaction with its previous steps and using several Q-tables, the method is more robust against adversarial users who have malicious aims in the system.

%The rest of this paper is organized as follows. Section \ref{section:LR} reviews related work, and we introduce the system model in Section \ref{section:Scenario}. Section \ref{section:proposed_method} includes the self-play formulation in the value-based problem. Section \ref{section:result} presents the simulation results, and Section \ref{section:conclusion} concludes this work. 

\section{Related Work}
\label{section:LR}

%To optimize resource allocation within network slices, ML-based methods have emerged as powerful tools. ML techniques, such as RL and DRL, enable intelligent decision-making for resource allocation by learning patterns from data and adapting to dynamic network conditions \cite{9333595}. ML algorithms can dynamically allocate resources, considering latency and throughput requirements \cite{zhou2022learning}. 

ML-based methods have emerged as powerful tools to optimize resource allocation within network slices, dynamically allocating resources while considering latency and throughput requirements \cite{zhou2022learning}. This adaptive approach enhances the efficiency and responsiveness of the network slicing, ensuring that resources are allocated optimally to meet various slice demands, and
%ML-based resource allocation in network slicing thus plays a pivotal role in realizing the full potential of 5G services, 
offering improved performance, reliability, and adaptability to different applications \cite{dangi2022ml}. Q-learning proves to be a valuable approach to resource allocation in network slicing and provides an adaptive and intelligent mechanism for allocating resources effectively \cite{zhou2022knowledge}.% \cite{elsayed2019aiQL}. %By leveraging Q-learning in network slicing, operators can strike a balance between the competing demands of eMBB and URLLC slices, optimizing resource utilization and enhancing 5G networks' overall performance for a wide range of applications.

While Q-learning is a powerful tool for 5G communications, 
%\notered{Why is Q-learning a powerful tool specifically for RB?}
it has several drawbacks such as the tendency to overestimate Q-values, exploration-exploitation trade-off, and slow learning in large state spaces, leading to suboptimal decision-making and slower convergence \cite{elsayed2019ai}. In \cite{zhou2022learning}, a DQN-based method is proposed for resource allocation to approximate Q-values, enabling the algorithm to handle more complex state-action spaces. %In \cite{rezazadeh2021zero}, a DQN is utilized to mitigate overestimation bias by maintaining two separate Q-tables, reducing suboptimal decisions. 
In \cite{wang2021adaptive}, an ensemble Q-learning is employed by multiple Q-learners, aggregating their predictions to improve overall performance and robustness. In \cite{peer2021ensemble}, an ensemble bootstrapped Q-learning method is proposed as a bias-reduced algorithm, extending double Q-learning to ensembles to mitigate both over-estimation and under-estimation biases, demonstrating superior performance in deep reinforcement learning (DRL). In \cite{wang2021adaptive}, an adaptive ensemble Q-learning method is proposed to address the overestimation issue in Q-learning by adjusting ensemble size based on upper and lower bounds of estimation bias, which demonstrates an improvement in model learning. While ensemble Q-learning enhances robustness and promotes continuous improvement, it encounters difficulties dealing with overestimation bias, leading to suboptimal decision-making. Additionally, the interdependence of Q-values across different models limits the efficient learning of multiple models simultaneously.
%While ensemble Q-learning enhances robustness and promotes continuous improvement, it has difficulties dealing with overestimation bias leads to suboptimal decision-making, and an interdependence of Q-values across different models limits the efficient learning of multiple models simultaneously. 
As a solution, self-play ensemble Q-learning is proposed, in which agents interact with themselves and the environment simultaneously by playing against previous versions of themselves, leading to learning and improvement action selection, rather than exploring random actions. %\notered{here you need to explain how self-play is different. you are merging self play with ensemble, What does self play add on top ensemble and why it is better. Just two sentences to differentiate your proposal from ensemble q-learning. I assume this hasnt been proposed before, right?} \notegreen{Ensemble Q-learning has been proposed, but self-play ensemble q-learning has not.} 

These advancements in Q-learning variants demonstrate a collective effort to overcome its limitations, offering more effective and adaptive resource allocation strategies in dynamic and complex environments of network slicing for 5G and beyond. While ML algorithms have considerable effects on improving the quality of services and meeting requirements, they have some issues related to the exploration-exploitation trade and they are also vulnerable to malicious users \cite{salehi2023adverserial}. Malicious users attempt to perturb the system and degrade its performance by altering the policy for resource allocation \cite{salehi2023adverserial} or introduce interference to the signal, leading to signal degradation in Q-learning scenarios \cite{salehi2024jamming}. The goal of this paper is to employ self-play ensemble Q-learning for the first time to enhance resource allocation performance in network slicing and improve robustness against adversarial users.% \notered{is the technique new or is the application to NS is new?clarify for me. make an internal comment}. %\notered{This is true, however many other algorithms have been used to solve this issue, thus, I would add here the adversarial part as well}. \notered{Would you be able to expand on how malicious users can affect non-malicious ones?} %\noteblue{ Then the impact of adversarial users on this system is assessed which aims to modify the system policy for resource allocation. %We evaluate the robustness of self-play ensemble Q-learning against a scenario influenced by a malicious user.}
%\notegreen{In this paper we proposed the self-play ensemble q-learning for the first time, and its application in network slicing is also novel. Furthermore, proposing a method for using self-play for the value-based method is also novel, as in previous papers, this method was considered in policy-based methods.}

\section{System Model}
\label{section:Scenario}

\subsection{System Scenario}

\begin{figure}[t!]
  \centering
  \includegraphics[width=1\linewidth]{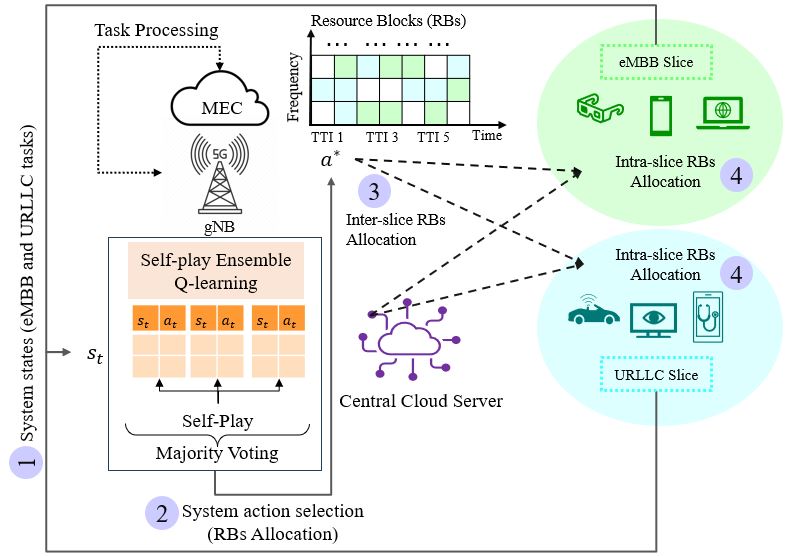}
  \caption{Network Slicing Scenario.
  }
    \label{fig:Sce}
    \vspace{-0.6cm}
\end{figure}

In this paper, a network-slicing scenario is considered for supporting two distinct types of slices, namely eMBB and URLLC slices \cite{zhou2022knowledge}. As illustrated in Fig. \ref{fig:Sce}, a two-step resource allocation scheme is employed on a gNodeB (eNB). The eNB is equipped with a mobile edge computing (MEC) server which provides the capability of offloading tasks to a cloud server. During the inter-slice phase, the eNB intelligently distributes radio resources among two slices. Subsequently, these allocated resources are utilized within each specific slice during the intra-slice phase. The primary objective is to assign radio resources which refers to time-frequency resource blocks between two slices to meet the latency and throughput requirements of URLLC and eMBB slices. 
For meeting the requirements of slices, decreasing the latency is pivotal which is evaluated by the equation below: 

\begin{equation}
      d = d^{Tx}+d^{rTx}+d^{que}+d^{edge},
      \label{eqn:delay_total}
\end{equation}
where $d^{Tx}$ is the transmission delay, $d^{rTx}$ is the re-transmission delay, $d^{que}$ is the queuing delay, and $d^{edge}$ is the processing delay in the MEC server.
%\noteblue{and $d^{Cloud}$ is the processing delay in the central cloud for computation tasks.}
$d^{Tx}$ is affected by the size of packets sent by a user equipment (UE) and the link capacity between UE $u$ and its connected eNB $j$, denoted as $C_{j,u}$. The value of $C_{j,u}$ is calculated using the equation below: % It must be noted that $d^{Cloud}$ is affected by downlink and uplink delay, and cloud queuing delay refers to the time it takes for the data to be processed by the cloud. 
    \vspace{-0.2cm}
{\footnotesize
\begin{multline}
    C_{j,u} = \sum_{r \in N_u} b_{RB} \log \Bigg(1+ \\
    \frac{p_{j,r}x_{j,u,r}q_{j,u,r}}{b^{RB}N_{0}+\sum_{j' \in J_{-j}}\sum_{u' \in u_{j'}}\sum_{r' \in N_{j'}}p_{j',r'} X_{j',u',r'} g_{j',u',r'}}\Bigg),
\end{multline}
}
where $N_u$ is the set of allocated resource blocks to the UE $u$, $b_{RB}$ is a RB's bandwidth, $N_{0}$ is noise power density, $p_{j,r}$ is the transmission power of RB $r$ in the eNB $j$. $x_{j,u,r}$ indicates a binary variable that illustrates whether RB $r$ is assigned to the UE $u$ or not, 
%\sout{$q_{j,u,r}$ refers to the channel gain of the eNB $j$ and UE $u$}
$q_{j,u,r}$ refers to the channel gain between eNB $j$ and UE $u$ over resource block $r$, $J_{-j}$ indicates the set of eNBs, except the $j^{th}$, $u_{j'}$ is the set of UEs in the eNB $j'$, and $N_{j'}$ is the set of total resource blocks in eNB $j'$. It is noteworthy to mention that computation decisions and task management functionalities can be deployed in the eNB or the non-real-time RAN Intelligent Controller from the O-RAN architecture. The readers are referred to \cite{polese2023empowering} for details of O-RAN architecture. %\noteorange{cite my ORAN paper with Tomasso.} 
To meet UEs requirements, the URLLC slice seeks to reduce latency, while the eMBB slice aims to increase throughput. Given these considerations, the allocation of resource blocks could be formulated as follows:

\begin{multline}
%\begin{aligned}
\max_{j} \quad w^{eMBB} \left( \frac{\sum_{u \in M^{eMBB}_{j}} B^{eMBB}_{j,u}}{|M^{eMBB}_{j}|} \right) + \\ w^{URLLC} \left(D^{tar} - \frac{\sum_{v \in M^{URLLC}_{j}} D^{URLLC}_{j,v}}{|M^{URLLC}_{j}|} \right),\\
%\end{aligned}
\label{eqn:reward}
\end{multline}

\begin{comment}
\begin{equation}
    \text{subject to} \quad C_{eMBB}^{j} + C_{URLLC}^{j} \leq C_{j}, \quad \forall j
\end{equation}

\begin{equation}
\begin{aligned}
& & \max_{j} \quad w^{eMBB}B^{eMBB,avg}_{j} + w^{URLLC}(D^{tar}-D^{URLLC,avg}_{j})\\
\end{aligned}
\label{eqn:reward}
\end{equation}

\begin{equation}
\tag{\themycounter3.a}
B^{eMBB,avg}_{j} = \frac{\sum_{u\in M^{eMBB}_{j}} B^{eMBB}_{j,u}}{|M^{eMBB}_{j}|}
\label{eqn:1a}
\end{equation}

\begin{equation}
\tag{\themycounter 3.b}
D^{URLLC,avg}_{j} = \frac{\sum_{v\in M^{URLLC}_{j}} D^{URLLC}_{j,v}}{|M^{URLLC}_{j}|}
\label{eqn:1b}
\end{equation}
\end{comment}

\begin{equation}
\tag{\arabic{mycounter}.a}
 \sum_{u\in M^{eMBB}_{j}} x_{j,u,r'} + \sum_{v\in M^{URLLC}_{j}} x_{j,v,r'} =1,
 \label{eqn:1c}
\end{equation}

\begin{equation}
\tag{\arabic{mycounter}.b}
\sum_{r' \in N_{j'}} ( \sum_{u\in M^{eMBB}_{j}} x_{j,u,r'} + \sum_{v\in M^{URLLC}_{j}} x_{j,v,r'} ),
\label{eqn:1d}
\end{equation}

\begin{equation}
\tag{\arabic{mycounter}.c}
C_{eMBB}^{j} + C_{URLLC}^{j} \leq C_{j},
\label{eqn:1e}
\end{equation}

where $w^{eMBB}$ and $w^{URLLC}$ are the weighting factors of eMBB and URLLC slices to make throughput and latency comparable. $B^{eMBB}_{j,u}$ represent the throughput of the UE $u$ in the eMBB slice of eNB $j$. $D^{tar}$ refers to the target delay of the URLLC slice, and $D^{URLLC}_{j,v}$ identifies the UE latency within the URLLC slice of eNB $j$. $M^{eMBB}_{j}$ and $M^{URLLC}_{j}$ are indicators of how many eMBB and URLLC slices are available in eNB $j$. Using eq. (3a) ensures that each RB is allocated to a single UE. According to eq. (3b) and eq. (3c), the total number of eNBs and computation capacity allocated to the eNB should not be greater than the number of resources available within the eNB.
\vspace{-0.2cm}

\section{Proposed Method}
\label{section:proposed_method}

The varying number of UEs belonging to URLLC and eMBB slices request access to radio resources. In the considered scenario, resource blocks are assigned in a centralized manner based on Q-learning, wherein the Q-table is updated using the equation below:  %\notered{Please, specify if this is centralized learning or you have distributed agents being each agent the eNB.}
    \vspace{-0.2cm}
\begin{multline}
      Q(s_t,a_t) = Q_{past}(s_t,a_t) + \\
      \alpha (r + \underset{a}{\gamma\max} Q(s_{t+1},a_{t+1})-Q_{past}(s_t,a_t)),
\end{multline}
where $s_{t+1}$ is the next state after taking action $a_t$ at state $s_t$, and $\alpha$ and $\gamma$ are the learning rate and discount factor, respectively. For using the Q-learning algorithm for RB allocation, the following components of the Markov decision process (MDP) are taken into account for the agent:

\begin{itemize}
\item {\bf States:} States for the agent is $(q^{eMBB}, q^{URLLC})$, indicating the number of eMBB and URLLC tasks in the queue. 

\item {\bf Action:} The agent has access to the radio resources, and its action is dictated by the number of resource blocks allocated to the eMBB and URLLC slices. The action set is defined by $(r^{eMBB}, r^{URLLC})$. 

\item {\bf Reward:} The reward function, as defined in Eq. \ref{eqn:reward}, is calculated based on the average throughput and delay experienced by the eMBB and URLLC slices. %\notered{Perhaps refer to eq. 3}
\end{itemize}

Q-learning involves updating the Q-values based on the maximum Q-value of the next state. Furthermore, Q-learning focuses on updating Q-values based on the agent's interactions with the environment, and the update rule considers the immediate reward obtained from taking an action in a given state and the estimated future rewards. However, this can result in an overestimation of Q-values, especially when the eNB has not sufficiently explored the environment. Furthermore, Q-learning has the challenges of hyperparameter fine-tuning. Foremost, adversarial agents have a significant and detrimental impact on Q-learning performance, leading to substantial degradation. For this reason, the self-play ensemble Q-learning algorithm has been suggested as a method for handling these challenges and improving the RB allocation.

\subsection{Self-play Ensemble Q-learning}
\label{subsection:SP_Ens}

The ensemble Q-learning approach involves maintaining several Q-tables, denoted as $Q_i$, where $i$ represents the number of tables. In the considered approach, decisions from multiple Q-tables are aggregated using a majority voting (MV) mechanism. Each agent provides its Q-values for a specific state-action pair, and the action with the majority of votes is selected as the final decision. $Q_i(s_t, a_t)$ represent the Q-value predicted by the $i$-th Q-learning agent for state $s_t$ and action $a_t$. The MV function can be defined as \textit{Algorithm 1} where $N$ is the number of Q-tables which is equal to 3, and ${a^*}$ is the selected action based on MV. While increasing the number of Q-tables improves the method's exploration capabilities, it also increases the method's complexity and energy consumption. Therefore, we select $N=3$ to balance these factors.

\begin{algorithm}[t!]
\caption{MV Function of Self-Play Ensemble Q-learning Method}
\SetAlgoLined
\For{Each state $s_t$}{
    Initialize a counter for each possible action $a_t$\;
    \For{$i$ to $N$}{
        Compute $Q_i(s_t, a_t)$ for all actions $a_t$\;
        Find the action $a_t^*$ with the highest Q-value $Q_i(s_t, a_t)$\;
        Increment the counter for action $a_t^*$\;
    }
    Select the action $a_t^*$ with the highest count as the final decision for state $s_t$\;
}
\end{algorithm}

%$\delta$ is the Kronecker delta function, 
When implementing MV for Q-learning, it's crucial to address ties when there is no clear majority. A random selection strategy is employed for this purpose. It must be noted that the weights of all considered $Q_i$ are assumed to be the same, then the system is affected by all of them equally. %It's important to mention that the weights of $Q_i$ votes are treated equally.

%where $N$ is the number of Q-learning agents, $\delta(x, y)$ is the Kronecker delta function (equal to 1 if $x = y$ and 0 otherwise), and ${a^*}$ is the selected action based on MV. When implementing MV for Q-learning, it's crucial to address ties when there is no clear majority. A random selection strategy is employed for this purpose.

Ensemble Q-learning, while effective in some scenarios, faces several challenges that can hinder its performance in dynamic network slicing environments. One of the challenges is its exploration, which increases the probability of convergence to local optima. To address this, self-play ensemble Q-learning considers its previous values, always striving to select the best action encountered thus far and avoiding getting stuck in local optima. Additionally, handling complex state spaces poses a difficulty for ensemble Q-learning. Self-play ensemble Q-learning introduces mechanisms to better understand and learn from all Q-values by the explicit consideration of the agent's interactions with its past versions. %This allows the agent to learn from its history and potentially improve its strategy over time through self-play experiences. 
Furthermore, the Q-values are updated not only based on the immediate rewards from the environment but also the outcomes of self-play experiences against its past versions. The update involves a weighted average of the current Q-values and the past versions of Q-values, which are calculated by the equation below.%}\notered{rewrite}:

\begin{equation}
%\vspace{-0.3cm}
    Q_i(s_{t+1}, a_{t+1})\leftarrow (1 - \beta)Q_i(s_t, a_t)+\beta Q_{past,i}(s_t, a_t),
    %\vspace{-0.3cm}
\end{equation}
where the weight $\beta$ determines how much influence the past version has on the current Q-values. This ensures a balance between incorporating new experiences from the current episode and leveraging knowledge from the past version. Using this equation leads to observing the performance of the current agent's observations from the environment against its previous strategies and controlling the influence of the past versions on the updating of current Q-values. This part of the algorithm focuses on enhancing the current agent's Q-values by considering its past strategies, allowing it to adapt and improve over time through self-play. For each $(s_{t+1}, a_{t+1})$, iterate through all $(s_t, a_t)$ in the Q-table. %In each iteration, simulate a game where the current agent (updated during the episode) plays against a past version of itself. This is essentially a self-play scenario where the agent assesses its performance against its historical strategies. Based on the simulated game, observe the outcomes, including rewards and transitions. In the self-play ensemble Q-learning algorithm, after each episode, there is an additional step that is updated against the past version. This step involves simulating a game against a past version of the agent and updating the Q-values based on the outcomes. This process allows the agent to learn from its past versions, potentially improving its performance over time through self-play experiences. So, the key difference is the incorporation of self-play experiences and updates against past versions, which aims to enhance the learning process through interactions with the agent's history. 

The self-play ensemble Q-learning method can improve the learning process in the following ways:
\begin{itemize}
    \item \textbf{Monitoring}: In the ensemble part, the agent attempts to consider various aspects of the system, choosing different actions in each state to receive diverse rewards. This process helps the agent generate diverse experiences.
    \item \textbf{Adaptability}: Self-play enables the agent to adapt its strategy over time. As it encounters different situations, it learns to respond to a variety of states, making it more robust and adaptable.
    \item \textbf{Continuous Improvement}: The agent improves its policies and strategies by repeatedly playing against its current or past versions, refining its decision-making process, discovering new tactics, and enhancing performance.
    \item \textbf{Reduced Dependency on External Data}: Since the agent generates its training data by self-play, it becomes less dependent on external data. This is useful when external data is poisoned, limited, or not readily available.
    \item \textbf{Exploration of Strategies}: During self-play, the agent explores various strategies, enhancing its understanding of the environment and potentially discovering optimal or near-optimal policies.
\end{itemize}

This process allows the current agent to learn from its recent experiences in the network slicing system, and also its history of playing against earlier versions. It introduces a form of memory or hindsight learning, potentially improving the agent's strategy over time. \textit{Algorithm 2} provides information about the self-play ensemble Q-learning on the RL algorithm. %The weighted average ensures a balance between new knowledge gained from recent experiences and the knowledge retained from the past.

\begin{comment}    
\begin{algorithm}[t!]
 \caption{Self-Play Ensemble Q-learning Method}
 \SetAlgoLined
 \textbf{MDP Parameters:} \(Q_i, s, a, r, \epsilon, \gamma, \alpha_i, \beta\)\\
 \For{t = 1 to TTI}{
    Initialize \(Q_i\) = 0\, and state \(s\)\;
    %Initialize state \(s\)\;
    \While{not end of episode}{
        Choose action \(a_t\) based on \(\epsilon\)-greedy policy from \(Q_i(s_t, \cdot)\)\;
        Take action \(a_t\), observe reward \(r_t\), and transition to next state \(s_{t+1}\)\;
        %\[ Q_i(s, a_t) \leftarrow (1-\alpha_i) Q_i(s_t, a_t) + \alpha_i (r_t + \gamma Q_{i}(s_{t+1}, a_{t+1})) \]  
        Q_i(s, a_t) \leftarrow (1-\alpha_i) Q_i(s_t, a_t) + 
              \alpha_i (r_t + \gamma Q_{i}(s_{t+1}, a_{t+1})) \\      
        \textit{Self-Play Update:}\;
        \For{each state \(s\)}{
            \For{each action \(a\)}{
                Simulate a game against a past version of the agent using \(Q_i\)\;
                Observe the outcomes and update \(Q_i\) using weighted averages:
                %\[ Q_i(s_{t+1}, a_{t+1}) \leftarrow (1 - \beta) Q_i(s_t, a_t) + \beta Q_{\text{past, i}}(s_t, a_t) \]
                Q_i(s_{t+1}, a_{t+1}) \leftarrow (1 - \beta) Q_i(s_t, a_t) + 
                           \beta Q_{\text{past, i}}(s_t, a_t)\\
            }
        }
        \textit{Majority Voting:}\;
        Choose the action with the majority votes among \(Q_i(s_t, \cdot)\) as the final action for state \(s\)\; 
        %Move to the next state \(s_{t+1}\)\;
    }
 }
\end{algorithm}
\end{comment}

\begin{algorithm}[t!]
 \caption{Self-Play Ensemble Q-learning Method}
 \SetAlgoLined
 \textbf{MDP Parameters:} \(Q_i, s, a, r, \epsilon, \gamma, \alpha_i, \beta\)\\
 \For{$t = 1$ \textbf{to} $TTI$}{
    Initialize \(Q_i(s, a) = 0\) and state \(s\)\;
    \While{not end of episode}{
        Choose action \(a_t\) based on \(\epsilon\)-greedy policy from \(Q_i(s_t, \cdot)\)\;
        Take action \(a_t\), observe reward \(r_t\), and transition to next state \(s_{t+1}\)\;
%        \[
%        Q_i(s_t, a_t) \leftarrow (1 - \alpha_i) Q_i(s_t, a_t) + \alpha_i \left(r_t + \gamma \max_{a} Q_i(s_{t+1}, a)\right)
%        \]
        \begin{align*}
        \begin{aligned}
        Q_i(s_t, a_t) & \leftarrow (1 - \alpha_i) Q_i(s_t, a_t) \\
        & + \alpha_i \left(r_t + \gamma \max_{a} Q_i(s_{t+1}, a)\right)
        \end{aligned}
        \end{align*}
        \textit{Self-Play Update:}\;
        \For{each state \(s\)}{
            \For{each action \(a\)}{
                Simulate a game against a past version of the agent using \(Q_i\)\;
                Observe the outcomes and update \(Q_i\) using weighted averages:
                \begin{align*}
                \begin{aligned}
                %\[
                Q_i(s_{t+1}, a_{t+1}) \leftarrow (1 - \beta) Q_i(s_t, a_t) \\
                + \beta Q_{\text{past}, i}(s_t, a_t)
                %\]
                \end{aligned}
                \end{align*}
            }
        }
        \textit{Majority Voting:}\;
        Choose the action with the majority votes among \(Q_i(s_t, \cdot)\) as the final action for state \(s_t\)\;
    }
 }
\end{algorithm}

\subsection{Adversarial Agent}
\label{subsection:Conf}
%\notered{the same comment as previous section. these two sections come very sudden}
In resource allocation for network slicing using RL algorithms, one of the Q-tables, denoted as $Q_i$, may be manipulated by a malicious user. The malicious user strategically alters the Q-values to introduce a deliberate bias in the learning process. The intentional manipulation introduced by the malicious user is denoted by the equation below: 

\begin{align}
\vspace{-0.3cm}
    \resizebox{0.90\linewidth}{!}{$
    Q_i(s,a) \leftarrow Q_i(s,a) + \alpha 
             \left( r + \gamma \min(Q_i(s',\cdot), Q_{i'}(s',\cdot)) - Q_i(s,a) \right),$}
             \vspace{-0.3cm}
\end{align}
in which malicious influence disrupts the integrity of the learning process. The algorithm, which aims to converge towards optimal policies based on accurate Q-value estimates, is misled by the distorted information provided by the adversarial agent. This adversarial interference can lead to suboptimal resource allocation decisions as the algorithm's learning dynamics are compromised. The impact is significant in dynamic environments where the learning agent relies on accurate Q-value estimations to adapt to changing network conditions. 
\vspace{-0.3cm}

\subsection{Baselines Learning Algorithms}
\label{subsection:Baselines}

In this paper, we considered two baselines, double Q-learning, and Q-learning, to evaluate the performance of the self-play ensemble Q-learning methods.

\subsubsection{Q-learning}
\label{subsubsection:QL}

We employ Q-learning, a foundational RL algorithm, as the baseline model for evaluating the effectiveness of the self-play ensemble Q-learning method. Q-learning, renowned for its robustness and versatility, serves as a benchmark in assessing the performance and advancements achieved by the proposed ensemble technique.

\subsubsection{Double Q-learning}
\label{subsubsection:DQL}
%\notered{where did this section come from? is this a baseline? or what? did you mention in the abstract and intro that you are having this as baseline.} \notegreen{To respond to this comment and the following two comments, I added more description to the Abstract and Introduction of the paper.} 
%\noteorange{ok, but now you have two baselines and you only talk about one, you should have this section as baselines and then make subtitles by underlining double Q and RL. RL can be one sentence but needs to be mentioned.}%\noteorange{Note: Perhaps in the subsection title should specify it is a baseline.Like Baselines: Double... However, I would say If it is the well-known double q learning without any modifications from the original paper, I think we should mention it in the simulation settings and reference its original paper. }
Double Q-learning is a model-free RL algorithm that uses two sets of Q-values, commonly referred to as $Q_A$ and $Q_B$. 
%As a basic principle, 
The selection of the best action should be decoupled from the estimation of its value. Rather than always using the same set of Q-values to determine the best action and to estimate its value, it uses one to select the best action and the other to estimate it. Double Q-learning reduces the overestimation bias associated with traditional Q-learning by using two sets of Q-values and updating them alternately.% \notered{This is a well-known algorithm, I don't know if we should keep the description in the text. For now, let's leave it and ask prof about it}. 

%In double Q-learning, after selecting the action, the corresponding Q-table is updated, in which \(Q\) is the Q-table used for action selection (either \(Q_A\) or \(Q_B\)). Double Q-learning, explicitly alternates between updating two sets of Q-values to address overestimation bias.
\vspace{-0.3cm}

\section{Simulation Results}
\label{section:result}

\subsection{Simulation Settings}

Considered scenario has three independent eNBs, with a cell radius of 125 meters and a bandwidth of 20 MHz, supporting 13 resource block groups. The network environment is based on the 3GPP Urban Macro network model. The eNBs operate with a PHY configuration that includes a 15 kHz subcarrier spacing, 12 subcarriers per resource block, and a transmission power of up to 40 dBm. The antenna gain is set at 15 dB, and the system operates at a carrier frequency of 30 GHz.

Each eNB includes URLLC and eMBB slices with 10 UEs and 5 UEs, respectively. Packet sizes are 50 bytes for URLLC and 100 bytes for eMBB, with traffic generated according to a Poisson distribution. The TTI size is 2 OFDM symbols, equating to 0.1429 ms. Hybrid automatic repeat request (HARQ) processes are asynchronous, featuring a round trip delay of 4 TTIs, 6 HARQ processes, and a maximum of 1 HARQ re-transmission.

Propagation characteristics include a path loss model defined as $128.1 + 37.6 \log_{10}(D[\text{Km}])$, Log-Normal Shadowing with an 8 dB standard deviation, a noise figure of 5 dB, and a penetration loss of 5 dB. In terms of learning algorithm parameters, $\alpha$, $\beta$, $\gamma$, and $\epsilon$ are set to 0.5, 0.5, 0.2, and 0.3, respectively. The specific learning rates for the self-play ensemble Q-learning, denoted as $\alpha_a$, $\alpha_b$, and $\alpha_c$, are 0.7, 0.8, and 0.9, respectively.

\subsection{Results}
\begin{figure}
    \centering
    \includegraphics[width=1\linewidth]{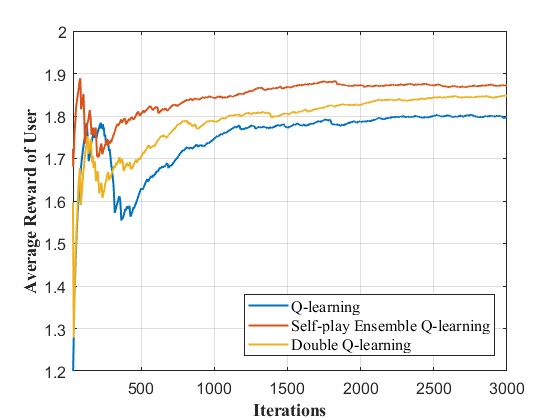}
    \caption{Average reward of the system}
    \label{fig:reward}
    \vspace{-0.7cm}
\end{figure}

%\notered{There is no mention of how much your algo is better than the baselines (percentages). You can find it on the abstract but anywhere else. I'm aware you are working on it, but remember to add a confidence interval to all figures.}
We evaluate the performance of three algorithms, namely Q-learning, double Q-learning, and self-play ensemble Q-learning in meeting the requirements of URLLC and eMBB slices. We then evaluate the robustness of self-play ensemble Q-learning and double Q-learning methods when one of the Q-tables is affected by a malicious user
%\notered{teh reader now sees why you have double q learning and adversarial, somethign like this should appear in intro and abstract}
. Fig. \ref{fig:reward} illustrates the convergence speed of the algorithms. The figure highlights that self-play ensemble Q-learning exhibits higher rewards compared to the other two algorithms. Moreover, this method achieves convergence to optimal rewards quicker than alternative approaches due to benefiting from multiple Q-value ensemble learning and competition with its past knowledge in self-play.

\begin{comment}
\begin{figure}
    \centering
    \includegraphics[width=0.9\linewidth]{Fig/Reward.eps}
    \caption{Average reward of the system}
    \label{fig:reward}
    \vspace{-0.7cm}
\end{figure}
\end{comment}

\begin{figure*}[!htb]
\minipage{0.33\textwidth}
  \includegraphics[width=\linewidth]{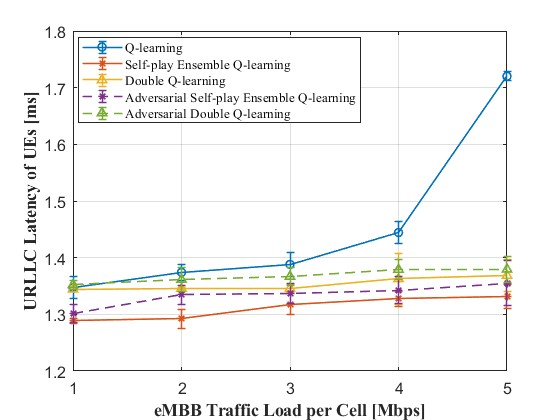}
  \caption{Latency of URLLC slices}\label{fig:URLLC}
\endminipage\hfill
\minipage{0.33\textwidth}
  \includegraphics[width=\linewidth]{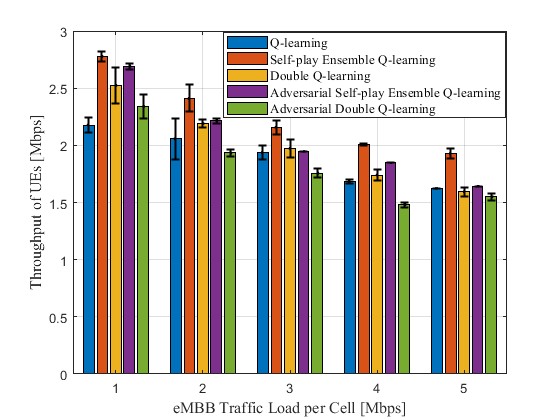}
  \caption{Throughput of eMBB slices}\label{fig:Throughput}
\endminipage\hfill
\minipage{0.33\textwidth}%
  \includegraphics[width=\linewidth]{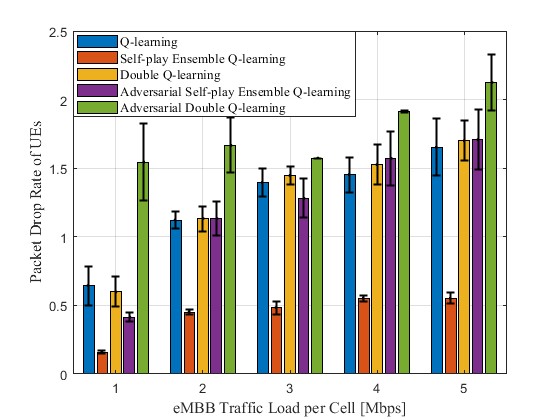}
  \caption{PDR of URLLC slices}\label{fig:PDR}
\endminipage
\end{figure*}

Fig. \ref{fig:URLLC} presents the latency of the URLLC slice when using Q-learning, double Q-learning, and self-play ensemble Q-learning. According to the results, the system delay with Q-learning is more significant than other algorithms, while self-play ensemble Q-learning exhibits the lowest delay. This method demonstrates 21.92\% improvement in latency. Similarly, as depicted in Fig. \ref{fig:Throughput}, self-play ensemble Q-learning exhibits a substantial enhancement, achieving 24.22\% improvement in throughput for the eMBB slice. Finally,  Fig. \ref{fig:PDR} showcases notable progress, boasting 23.63\% improvement in the packet drop rate (PDR) of self-play ensemble Q-learning is considerably lower than Q-learning and double Q-learning.

\begin{comment}  
\begin{figure*}[ht]
  \subfloat[Latency of URLLC slices]{
	\begin{minipage}[c][0.7\width]{
	   0.32\textwidth}
	   \centering
	   \includegraphics[width=1.1\textwidth]{Fig/URLLC.eps}
        \label{fig:URLLC}
	\end{minipage}}
 \hfill 	
  \subfloat[Throughput of eMBB slices]{
	\begin{minipage}[c][0.7\width]{
	   0.32\textwidth}
	   \centering
	   \includegraphics[width=1.1\textwidth]{Fig/Throughput.eps}
        \label{fig:Throughput}
	\end{minipage}}
 \hfill	
  \subfloat[Packet drop rate of URLLC slices]{
	\begin{minipage}[c][0.7\width]{
	   0.32\textwidth}
	   \centering
	   \includegraphics[width=1.1\textwidth]{Fig/PDR.eps}
        \label{fig:PDR}
	\end{minipage}}
\caption{System performance under various RL-based algorithms.}
\end{figure*}
\end{comment}

\begin{comment}    
\begin{figure*}[!htb]
\minipage{0.33\textwidth}
  \includegraphics[width=\linewidth]{Fig/URLLC.eps}
  \caption{Latency of URLLC slices}\label{fig:URLLC}
\endminipage\hfill
\minipage{0.33\textwidth}
  \includegraphics[width=\linewidth]{Fig/Throughput.eps}
  \caption{Throughput of eMBB slices}\label{fig:Throughput}
\endminipage\hfill
\minipage{0.33\textwidth}%
  \includegraphics[width=\linewidth]{Fig/PDR.eps}
  \caption{PDR of URLLC slices}\label{fig:PDR}
\endminipage
\end{figure*}
\end{comment}

\begin{comment}    
\begin{figure*}
\centering
\subfloat[Latency of URLLC slices]{\label{fig:URLLC} \includegraphics[width=0.33\textwidth]{Fig/URLLC.eps}}% The "%" masks the line break.
\hfill
\hspace{-3em} % Adjust the space between the figures
\subfloat[Throughput of eMBB slices]{\label{fig:Throughput} \includegraphics[width=0.33\textwidth]{Fig/Throughput.eps}}%
\hfill
\hspace{-3em} % Adjust the space between the figures
\subfloat[Packet drop rate of URLLC slices]{\label{fig:PDR} \includegraphics[width=0.33\textwidth]{Fig/PDR.eps}}%
\caption{System performance under various RL-based algorithms.}
\label{Expert_Le}
\end{figure*}
\end{comment}

In the context of a scenario where one of the tables, is affected by a malicious user, as depicted in Fig. \ref{fig:URLLC} to Fig. \ref{fig:PDR}, the adversarial agent intentionally manipulates Q-values to mislead the resource allocation decisions of the learning algorithm. Such manipulation can significantly impact the overall performance of the system. It must be noted that double Q-learning relies on two separate Q-tables to estimate action values, the malicious agent's influence directly skews the learning process and can misguide the algorithm's decision-making. Then the results of resource allocation become suboptimal and deviate from the system's desired objectives. In self-play ensemble Q-learning, degradation of the system performance occurs when the affected Q-value misleads the system with intentionally biased information. Self-play ensemble Q-learning is less affected by malicious users than double Q-learning because it uses multiple Q-tables to cross-check and detect irregular values, helping the agent avoid actions influenced by malicious behavior. This approach improves robustness by leveraging historical Q-values to identify and correct deviations, ensuring more stable learning and decision-making processes in the presence of adversarial interference. %However the impact of that malicious user on self-play ensemble Q-learning is significantly lower than on double Q-learning, as the agent, by considering the past Q-values of Q-tables, is better able to distinguish irregular values and avoid selecting those actions.

\section{Conclusion}
\label{section:conclusion}

In this paper, we propose a novel self-play ensemble Q-learning approach for resource allocation in a network slicing scenario, aiming to meet the requirements of two different slices. With self-play ensemble Q-learning, the system utilizes the Q-values from three Q-tables and incorporates the agent's knowledge from previous steps. This method successfully fulfills a range of requirements, achieving improvements of 21.92\% in latency, 24.22\% in throughput, and 23.63\% in PDR compared to Q-learning. Additionally, through the voting of three different agents for action selection, the method demonstrates greater robustness against malicious users.%\notered{Please, refer to the gains of Self-play Ensemble Q-Learning over the baselines.} 

%\notered{Fix the reference format. Authors, remove underscore, and use abbreviations for Journals and conferences. }

\section*{Acknowledgment}

This work has been supported by MITACS and Ericsson Canada Inc., and NSERC Canada Research Chairs Program.

%\bibliographystyle{plain}  
%\bibliography{ref} 

\end{document}